\documentclass[prd,eqsecnum,showpacs,aps,nofootinbib]{revtex4}
\usepackage{latexsym}
\usepackage{graphicx}
\begin{document}
\title{On astrophysical explanations due to cosmological inhomogeneities
\\ for the observational acceleration}
\author{Kenji Tomita }
\affiliation{Yukawa Institute for Theoretical Physics, 
Kyoto University, Kyoto 606-8502, Japan}
\date{\today}

\begin{abstract}
We review various cosmological models with a local underdense region
(local void) and the averaged models with the backreaction of
inhomogeneities, which  
have been proposed to explain (without assuming a positive cosmological
constant) the observed accelerating behaviors appearing in the
magnitude-redshift relation of SNIa. To clarify their reality, we
consider their consistency with the other observational studies such
as CMB temperature anisotropy, baryon acoustic oscillation, kinematic
Sunyaev-Zeldovich effect, and so on. It is found as a result that
many inhomogeneous models seem to be ruled out and only models with
the parameters in the narrow range remain to be examined, and
that, unless we assume very high amplitudes of perturbations or  
gravitational energies, the averaged models cannot have the
accelerated expansion and the fitted effective $\Lambda$ has not the 
value necessary for the observed acceleration.
\end{abstract}
\pacs{98.80.-k, 98.70.Vc, 04.25.Nx}

\maketitle


\section{Introduction}
\label{sec:level1}
The observational magnitude ($m$) - redshift ($z$) relation for
distant supernovas of type Ia (SNIa) have been studied by the High-$z$
SN Search Team\cite{schm,ries1,ries2} and the Supernova Cosmology
Project\cite{perl}, and among the Friedman-Lemaitre-Robertson-Walker
cosmological models, the $\Lambda$-dominated models have been selected
as the best models which can explain their accelerating behavior. At
present, the model parameters $(\Omega_M, \Omega_{\Lambda}) \simeq
(0.25, 0.75)$ are regarded as the representative one.

About 40 years ago, the cosmological constant havs been paid attention
to avoid the contradiction between the age of oldest stars and the
cosmic age\cite{tomh,stab,refs,solh}, but its value has been 
uncertain, since the Hubble constant 
was inaccurate. But recently the existence of $\Lambda$ has been
realistic and reduced to the above value.

The $\Lambda$-dominated models with the above parameters have recently
been found to be consistent with all results of many observations such
as CMB temperature anisotropies\cite{map,spg}, the baryon acoustic
oscillation (BAO)\cite{eisens,seo,perc1,perc2,perc3}, statistics of
cluster abundance\cite{bahc1,bahc2,wang} and the inverse Sachs-Wolfe
effect\cite{granett,tomino}, and so it is regarded as a concordant
model.   

In these models, however, there are well-known unsolved problems: the
cosmological-constant problem and the incidence problem. The former
problem shows that the standard value of the cosmological constant is
too small ($\sim 10^{-120}$), compared with a constant appearing as
the zero-pont fluctuation of a field. The latter shows that the role
of the cosmological constant is important only at the later stage of $z <
1$, when structure formations are remarkable. Moreover we know the
low-$l$ anomaly in the CMB temperature anisotropies which remain to be
unsolved\cite{olv,cont}. 

In view of these problems, on the other hand, astrophysical
explanations for the accelerating property of SNIa were tried by
considering the optical role of inhomogeneous matter distribution
which may cause the apparent cosmological acceleration without
$\Lambda$ or dark energy. The explanations are divided into the
following two types by what inhomogeneity is assumed:

\noindent (1) {\it Non-Copernican inhomogeneity.} We assume to live in
a special point of the zero $\Lambda$ universe with a spherically 
symmetric underdense local inhomogeneity, which is called a local
void.  The apparent accelerating behavior of SNIa is 
obtained by deriving the $m$-$z$ relation in this inhomogeneous model.
       
\noindent (2) {\it Copernican inhomogeneity.} The universe is assumed
to have an uniform distribution of density perturbations with zero
$\Lambda$. The accelerating behavior is obtained by deriving the 
effective $\Lambda$ from them in the two procedures: 

\noindent a)\ Averaging and backreaction of inhomogeneous models with
$\Lambda = 0$. First the average model with the effective $\Lambda$
is derived and the $m$-$z$ relation is obtained  to to examine the
accelerating behavior.

\noindent b)\ Fitting.  By comparing the $m$-$z$ relation in a
perturbed model with zero $\Lambda$ and that in an unperturbed model
with nonzero $\Lambda$, the best fitted value of effective $\Lambda$
is obtained.

The excelent review about the works on (1) and (2)a), b) was
written by C${\rm \acute{e}}$l${\rm \acute{e}}$rier\cite{celerv} in
2007. In this note we have a review of the works which include
more recent theoretical and observational contributions. 

\section{A local inhomogeneity and SNIa}
\label{sec:level2}
Soon after the contributions by the High-$z$ SN Search
Team\cite{schm,ries1,ries2} and the Supernova Cosmology
Project\cite{perl}, another possibility was suggested independently by
 C${\rm \acute{e}}$l${\rm \acute{e}}$rier\cite{cele}, Goodwin et
al. (unpublished)\cite{good}, and me\cite{toma,tomb}. It was that 
the $m$-$z$ relation of SNIa
may be reproduced also in {\it non-Copernican} inhomogeneous cosmological 
models with a central underdense region. 

C${\rm \acute{e}}$l${\rm \acute{e}}$rier\cite{cele} discussed the
accelerated expansion from the qualitative viewpoint with  
general inhomogeneous models, and Goodwin et al.\cite{good}
discussed at the
similar epoch with a physical analysis of SNIa result due to the local
to global Hubble-constant ratio.

In my first model\cite{toma,tomb}, the inhomogeneity is described by
an open FRW solution  (in the inner underdense region) and a self-similar
solution (in the outer overdense region). The latter is a special case
of Lemaitre-Tolman-Bondi (LTB) solutions which is connected with the
former open FRW solution at a boundary, in order to avoid the central
singularity, and tends to the Einstein-de Sitter (EdS) solution in the
limit of radial infinity.  It was found in this model that the
theoretical $m$-$z$ relation can be consistent with the observed
relation, by adjusting the Hubble constants ($H_0$) and the total
density parameters ($\Omega_{M0}$) in the two regions, in the case
when the boundary is at the distance of $z \sim 0.07$ from the
center. The super-horizon version of this model, in which the distance
of the boundary is larger than the horizon size, was once proposed in
1995\cite{tomsp}.      

In my second model\cite{tomc}, two FRW models with $\Lambda = 0$
are used 
to build an inhomogeneous model, that is, an open FRW model (in the
inner underdense region) and a flat FRW model (in the outer overdense
region) connected with a discontinuous, sharp boundary with $z \approx
0.07$. The angular diameter distance $d_A$ is derived, solving the
Dyer-Roeder equation, the theoretical $m$-$z$ relation was derived,
and it was shown that the observed $m$-$z$ relation can be reproduced
similarly by adjusting the values of $H_0$ and $\Omega_{M0}$ in the
two regions. 

Moreover, in my next paper\cite{tomd}, the quantitative comparison
between the observed and theoretical $m$-$z$ relations was done using
the second inhomogeneous model and the best model parameters were
estimated.

Subsequently, Iguchi, Nakamura and Nakao\cite{igu} analyzed general
inhomogeneous models using LTB solutions. They reproduced the
$m$-$z$ relation in the concordant model, and it 
was shown that, at an intermediate radius (with $z < 1.7$), a critical
point appears in the models when we assume the uniform big-bang time and 
asymptotic vanishing spatial curvature.

The geometrical structure of LTB models mimicking $\Lambda$ or dark
energy was analyzed by Vanderveld et al.\cite{vand1} in detail and it was
shown that a weak central singularity and a critical point appear
generally. Recently Yoo, Kai and Nakao\cite{yoo} have showed that the
LTB models, which reproduce the $m$-$z$ relation in the concordant model
and have no critical point,  
can be obtained only under the condition of uniform big-bang time,
by improving Iguchi et al.'s model, and Clifton et al.\cite{clif}
derived LTB models without central weak singularity and critical point
and discussed the 
reproduction of the observed $m$-$z$ relation in their model.

The works on reproducing the observed $m$-$z$ relation in LTB models
were done successively by Alnes, Amarzguioui and Gr${\rm
\o}$n\cite{aag1,aag2}, Alnes 
and Amarzguioui\cite{aa,aadp}, Mansouri\cite{mans1,mans2},
Moffat\cite{moff1,moff2}, Biswas et al.\cite{bis}, and Alexander et
al.\cite{alex}.  

On the other hand, Kasai\cite{kasai} found from SN data themselves
@that they can be
divided into the low $z$ group (with $z < 0.2$) and the high $z$ group
(with $z > 0.3$), which correspond to higher and 
lower Hubble constants, and that the different trend of the data 
with respect to redshifts may represent the inhomogeneity of cosmological 
models.

\section{Consistency of inhomogeneous local-void models with the other 
observations}
\label{sec:level3}
In order that the inhomogeneous local-void cosmological models may be
realistic, they must be consistent with 
not only SNIa data, but also the other observations such as CMB
temperature anisotropies, BAO, the kinematic Sunyaev-Zeldovich effect,
and so on. In the following, let us review the works investigating 
the consistency with these observations.

\subsection{CMB temperature anisotropies}
We can see several acoustic peaks in the correlation ($C_l$) -
multipole ($l$) diagram for $l >200$, given with the WMAP data.
Since the inhomogeneous models are equal or nearly equal to the
Einstein-de Sitter (EdS) model at the recombination epoch, the EdS
model must be consistent with the observed property of CMB anisotropies.
First, Alnes et al.\cite{aag1} discussed the consistency for the first
peak 
and Alexander et al.\cite{alex} showed that their model (the Minimum
model) could give the consistent first and second peaks by assuming a
value of the Hubble 
constant (in the outer region) significantly lower than the
conventionally accepted value ($70$). The consistency with CMB was
discussed also by Blanchard et al.\cite{blan} and Hund and
Sarkar\cite{hund} in small values 
of the Hubble constant ($\sim 47$) in the outer region.
       
If an observer is off-center, he observes the dipole component of CMB
temperature anisotropies, proportional to the distance $r$ from the
center of the models. So this value $r$ is constrained by the observed upper
limit of the dipole anisotropy and the upper limit of $r$ is about 15 Mpc
\cite{tomdip,moff1,aadp}.    

\subsection{BAO}
The above acoustic peaks (representing an oscillation of photon-baryon
fluid around and before the recombination epoch) appear also as the baryon
acoustic oscillation (BAO), which is an oscillation of baryons (at the
later stage) imprinted in the matter spectrum.

Recently it was detected in the SDSS and 2dFGRS surveys, and the BAO
scale has been regarded as a standard
ruler\cite{eisens,seo,perc1,perc2,perc3}.  The characteristic (BAO)
scale is the sound horizon at the recombination epoch :
\begin{equation}
  \label{eq:b1}
r_s (z_{rec}) = \int_{z_{rec}}^\infty \ dz \ C_s (z) / H(z),
\end{equation}
where $C_s (z)$ is the sound speed at redshift $z$ and $z_{rec}$ is
the redshift at the recombination epoch. This scale is approximately
expressed as
\begin{equation}
  \label{eq:b2}
r_s \approx 147 (\Omega_{0} h^2/0.13)^{-0.5} (\Omega_{b}
h^2/0.024)^{-0.08} \ {\rm Mpc},
\end{equation}
where the density and Hubble parameters have the values at the place
we notice.

The observed scales of BAO brought from the galaxy samples in the
above surveys are used to constrain the values of the distance measure
$d_V (z) \equiv [(1+z)^2 (d_A)^2 cz/H(z)]^{1/3}$. Here $d_A$ is the
angular diameter distance and $H(z)$ is the Hubble parameter at redshift
$z$, and $d_V$ is the average of the scales perpendicular and parallel
to the line-of-sight directions. Percival et al.\cite{perc1} have recently
obtained the following relations at $z = 0.2$ and $0.35$:
\begin{eqnarray}
  \label{eq:b3}
r_s/d_V(0.2) &=& 0.1980 \pm 0.0058 \cr 
r_s/d_V(0.35) &=& 0.1094 \pm 0.0033. 
\end{eqnarray}
In the $\Lambda$ dominated concordant models, $r_s/d_V(0.2)$ and
$r_s/d_V(0.35)$ can be reproduced approximately (in about $95\%$).  In
my second models with a local void, on the other hand, these two
ratios are about $80 \%$ of the observed values. This is because the
epochs $z = 0.2$ and $0.35$ belong to the outer region with the EdS
model. So my second models cannot reproduce 
them, and they are therefore ruled out. In my first model and
the Minimum model given by Alexander et al.\cite{alex} also, the 
situation is similar and they are ruled out, as long as the boundary is
of the order of $300$ Mpc.  

In order that inhomogeneous models may not contradict with the BAO
observation, the epoch $z = 0.35$ must belong to the inner underdense
region, so that the scale of the inhomogeneous models must be Gpc
size. From this viewpoint, the Gpc-size inhomogeneous (LTB) models
have recently been studied by Clifton et al.\cite{clif} and
Garc\'ia-Bellido and Haugb${\rm \o}$lle\cite{bellido1}.

It has recently been found that the scale of BAO in the parallel to
line-of-sight or in the radial direction (the radial BAO scale)
impose on the models a more stringent condition than the BAO in the
perpendicular direction. RBAO has been studied by Gazta\~naga et
al.\cite{gazt} and it was found that the concordant models are
consistent with 
the observational result of RBAO. On the other hand, Zibin et
al.\cite{zibin} showed that RBAO imposes stringent conditions to Gpc-size
inhomogeneous models, using Gazta\~naga etal.'s data at $z = 0.24$ and
$0.43$. They considered the two types of models: a constrained model and
an unconstrained model, which were made so as to reproduce the recent
SN data and WMAP data with the boundary of $z \approx 1$, and
investigated the consistency with RBAO for the above
two redshifts. The constraint condition is $\int \delta \rho (t_i) r^2
dr \leq 0$. Their result is 

\noindent (1) \ the unconstrained model is consistent with the RBAO
data, but $H_0$ (the 
present Hubble constant) must be $\sim 44$, so that this model is
ruled out, and

\noindent (2) \ the constrained model is approximately consistent with
the RBAO data and $H_0$ is $\sim 60$, so that this model is not ruled
out at present. However the consistency is not so good as the
concordant models. 

\subsection{Kinematic Sunyaev-Zeldovich effect}   
The Sunyaev-Zeldovich (SZ) effect is a small spectral distortion of
the CMB radiation caused by the collisions of CMB photons with hot
thermal electrons. CMB photons passing through the hot
center of massive clusters interact with their electrons and take a
small distortion in the CMB spectrum due to the inverse-Compton
scattering. 

If the clusters are moving with respect to the CMB rest frame, there
is an additional spectral distortion (the kinematic Sunyaev-Zeldovich
effect) due to the Doppler effect of the cluster velocities on the
scattered CMB photons. If $v_{pec}$ is the component of the cluster
velocity along the line of sight, then the Doppler effect leads to the
following distortion of CMB spectrum:
\begin{equation}
  \label{eq:b4}
{\Delta T_{SZ} \over T_{CMB}} = \tau_e (v_{pec}/ c),
\end{equation}
where $\tau_e \equiv n_e \sigma_T R$ \ ($n_e, \sigma_T, R$ are the
electron density, the Thompson scattering cross-section and the
effective radius of a cluster). 

In inhomogeneous models with a local void, the CMB photon received by
a central observer is emitted at the recombination epoch in the outer
EdS region, in which the Hubble constant $(H_0)_{eds}$ is $\approx 47$
km/s/Mpc. In the regions inside and in the neighbourhood of the local
void, the Hubble constant $(H_0)_{loc}$ is assumed to be larger than
$(H_0)_{eds}$. So a cluster in the distance $r$ from the center of the
inhomogeneous models has a velocity $[(H_0)_{loc} - (H_0)_{eds}] r$,
relative to the CMB rest frame. This velocity can be observed as a
peculiar velocity ($v_{pec}$) of clusters in the kinematic
Sunyaev-Zeldovich effect. 

The possibility to observe the cluster
velocities systematically and put a constraint on cosmological models
was studied by Benson et al.\cite{ben}. 
Recently, Garc\'ia-Bellido and Haugb${\rm \o}$lle\cite{bellido2}
have shown the constraint  
due to the kinematic Sunyaev-Zeldovich effect using observed 9 clusters, and
shown that a strong constraint is given to the Gpc LTB models. It is
found that the models with the local void region larger than $\sim
1.5$ Gpc are ruled out practically, and only special Gpc LTB models
with a limited range of cosmological parameters may be allowed. 
But, if a systematic peculiar motion of clusters is discovered, this
effect may support the local-void model. 

\subsection{Spectral distortion of CMB radiation in the reionized
region}
In inhomogeneous models with a local void, the inner void region is
considered to be at the reionized stage. The ionized gas there is moving
outward, relative to the CMB frame, and leads to the Doppler
effect. When CMB photons are reflected 
by this ionized gas and reach an observer at the center of the void
region, spectral distortion appears in the accepted CMB radiation
owing to the Doppler anisotropy.
Here the reionized matter plays a role of moving mirror.
Caldwell and Stebbins\cite{cald} derived this distortion in the
inhomogeneous models with a local void of various sizes and showed that the
constraints for the models can be obtained from the observed upper
limit to the distortion of CMB spectrum. As a result, they found that
severe constraints can be obtained and the models with the local void of
the largest size ($\sim 2.5$ Gpc) are ruled out.

\section{Uniform distributions of density perturbations}
The appearance of observational acceleration due to the {\it
Copernican} uniform 
distribution of density perturbations has been studied by the
following two different forms.

\subsection{Averaging and backreaction of inhomogeneous models}
Averaging and backreaction have been studied by many workers,
who include Buchert\cite{buch}, Buchert and Corfora\cite{bcorf}, Ellis
and Buchert\cite{ell}, Kolb et al.\cite{kolb1,kolb2,kolb3}, Kolb et
al.\cite{kolb4}, Kasai\cite{kas1,kas2,kas3} and
Nambu\cite{nam1,nam2,nam3,nam4} for various gauges, and
Zalaletdinov\cite{zalal} and 
Paranjape\cite{paran} for a covariant form.

In the Buchert formalism, we consider an inhomogeneous model with dust,
and its metric is expressed in the comoving and synchronous gauge as
\begin{equation}
  \label{eq:d1}
ds^2 = -dt^2 + q_{ij} (t, x^m) dx^i dx^j.
\end{equation}
For averaging, we specify a hypersurface $\Sigma$ of constant $t$, and
take a compact region $D$ of $\Sigma$. If the volume of $D$ is $V_D$
and $\psi$ is a scalar variable, the average $<\psi>_D$ of $\psi$ over
$D$ is 
\begin{equation}
  \label{eq:d2}
<\psi>_D \equiv {1 \over V_D} \int_D \psi d\Sigma.
\end{equation}
Especially, for the matter density we have
\begin{equation}
  \label{eq:d3}
<\rho>_D \equiv {1 \over V_D} \int_D \rho d\Sigma.
\end{equation}
The averaged scale factor $a_D$ is defined by
\begin{equation}
  \label{eq:d4}
a_D \equiv (V_D)^{1/3}.
\end{equation}
Then we obtain from the Einstein equation

\begin{eqnarray}
  \label{eq:d5}
3{\ddot{a}_D \over a_D} &=& -{\kappa^2 \over 2} <\rho>_D + Q_D, \cr
3\Bigl({\dot{a}_D \over a_D}\Bigr)^2 &=& \kappa^2 <\rho>_D -{1\over 2}
<{\cal R}>_D - {1\over 2} Q_D 
\end{eqnarray}
and 
\begin{equation}
  \label{eq:d6}
(a_D^6 Q_D)^. + a_D^4 (a_D^2 <{\cal R}>_D )^. = 0,
\end{equation}
where ${\cal R}$ is the scalar curvature of $\Sigma$, 
\begin{equation}
  \label{eq:d7}
Q_D \equiv {2\over 3} (<\theta^2>_D - <\theta>^2_D) - (\sigma_{ij}
\sigma^{ij})_D, 
\end{equation}
and $\theta$ is the expansion of the world line of the dust fluid. In
the derivation of these equations, the incommutability between the
time derivative and the averaging was used:
\begin{equation}
  \label{eq:d8}
<\psi>_D^. = <\dot{\psi}>_D + <\theta\psi>_D - <\theta>_D<\psi>_D,
\end{equation}
where $<\dot{\psi}>_D \equiv <\partial\psi/\partial t>_D$ and
$<\psi>_D^. \equiv \partial <\psi>_D/\partial t$. The condition of
averaged acceleration $\ddot{a}_D > 0$ is given by
\begin{equation}
  \label{eq:d9}
Q_D > {1\over 2} \kappa^2 <\rho>_D.
\end{equation}
If this condition is satisfied, it seems that the averaged acceleration
may be realized, and so it has been studied by many workers under what
situation it is satisfied. 

Nambu and Tanimoto\cite{namtani} considered the region $D$ which
consists of many homogeneous and isotropic small regions with
different scale factors. Then it was shown that, even if we have
deceleration in each region, the average scale factor $a_D$ can have
acceleration such as $\ddot{a}_D > 0$. That is, the deceleration
observed in each region can be compatible with the averaged
acceleration. Ishibashi and Wald\cite{ishi} discussed this situation
and found that the Buchert averaging procedure has ambiguity both with
regard to 
the choice of time slicing and the choice of domain $D$, and that their
special choices can artificially derive the averaged acceleration.

Independently of Nambu and Tanimoto, \ R${\rm \ddot{a}}$s${\rm
\ddot{a}}$nen\cite{ras} 
considered a model with two disjoint regions consisting of an
overdense region and a completely empty region. In both regions we
have separately FRW spacetimes with scale factors $a_1$ and $a_2$,
where $a_1 \propto (1 - \cos u)$ and $t \propto (u - \sin u)$ and $a_2
\propto t$. If we define the averaged scale factor $a$ by $a^3 \equiv
{a_1}^3 + {a_2}^3$, then the averaged deceleration parameter $q \equiv (
\ddot{a}a)/\dot{a}^2$\  becomes negative and so acclerating, when the
overdense region turns around and starts collapse. However, 
R${\rm \ddot{a}}$s${\rm \ddot{a}}$nen's model is physically incomplete,
because the junction condition is neither used nor satisfied. 
Paranjape and Singh\cite{psing}, on the other hand, derived a LTB solution
corresponding to R${\rm \ddot{a}}$s${\rm \ddot{a}}$nen's model, in 
which there are three regions corresponding to R${\rm \ddot{a}}$s${\rm
\ddot{a}}$nen's two regions 
and a medium region. As a result, they found that $q < 0$ cannot be
realized and so we have not obtained the averaged acceleration in the
physically adjusted solution.

Next, Ishibashi and Wald\cite{ishi} insisted that in the Newtonian 
gauge the perturbed universes can be expressed by the metric 
\begin{equation}
  \label{eq:d10}
ds^2 = -(1 + 2\Psi) dt^2 + a^2(t) (1 -2\Psi) \gamma_{ij} dx^i dx^j,
\end{equation}
where $\gamma_{ij}$ denotes the metric of constant-curvature space.
The potential $\Psi$ is related to the matter density by the
cosmological Poisson equation
\begin{equation}
  \label{eq:d11}
a^{-2} \Delta_{(3)} \Psi = {1\over 2}\kappa^2 \delta \rho = {1\over
2}\kappa^2 (\rho - \bar{\rho}),
\end{equation}
where $\Delta_{(3)} \equiv \gamma^{ij} D_i D_j$ and $D_i$ denotes the
derivative corresponding to $\gamma_{ij}$. Here $\Psi$ 
satifies $|\Psi| <<1, \ |\partial \Psi/\partial t|^2 << a^{-2}
D^i\Psi D_i \Psi,$ and $(D^i\Psi D_i\Psi)^2 << (D^iD^j) D_iD_j \Psi$.
In this expression of spacetimes, the deviation of perturbed universes
from the FLRW spacetime is very small, even if there are nonlinear
perturbations like galaxies, clusters, voids and superclusters.
This is because $\Psi \sim G\delta M/(c^2 R) << 1$ for such
perturbations, where $\delta M \approx \delta \rho R^3$ and $R$ is the
radii of these structures.

The above analyses in the Newtonian gauge have a long history. Nariai
and Ueno\cite{nariai} and Irvine\cite{irv} derived cosmological
equations in the cosmological Newtonian approximation, and they have
been applied to the nonlinear treatment of matter evolution and the
N-body simulation.  Subsequently
the formulation in the cosmological post-Newtonian approximation also
was treated\cite{tompn,futam,shibata,matt}. 

From the viewpoint of Newtonian approximation, it is expected that the
averaging and backreaction of sub-horizon perturbations have only very
small contributions to the background. Kasai et al.\cite{kaf} derived
the no-go 
theorem in the Newtonian gauge that the nonlinear backreaction neither
accelerates nor decelerates the cosmic expansion, because the cosmic
averaged acceleration $\ddot{a}/a$ is determined merely by the mean
density as
\begin{equation}
  \label{eq:d11}
{\ddot{a}\over a} = - {4\pi G \over 3} <\rho>
\end{equation}
and the nonlinear backreaction reduces the expansion rate $\dot{a}/a$
as
\begin{equation}
  \label{eq:d12}
\Bigl({\dot{a}\over a}\Bigr)^2 = {8\pi G \over 3} <\rho> - {1\over 9a^2}
<D^i\Psi D_i \Psi> 
\end{equation}
Paranjape and Singh\cite{psing} and Siegel and Fry\cite{siegel} also
analyzed the possibility of averaged 
expansion in the Newtonian gauge and obtained the  negative results.

In the super-horizon case, Kolb et al.\cite{kolb1,kolb3} and Barausse et
al.\cite{barau} studied the possibility of averaged acceleration which
is caused by second-order super-horizon perturbations associated with
primordial inflation, but Flanagan\cite{flan} and Geshnizjani et 
al.\cite{gesh} obtained the negative results
because of the incompleteness in their second-order analysis and the
constraints from the CMB anisotropies. 

On the other hand, Kai et al.\cite{kai} studied the possibility of 
accelerated expansion by use of the LTB solution and found that, in a
strong inhomogeneous condition such as in R${\rm \ddot{a}}$s${\rm
\ddot{a}}$nen's one, the
acceleration of cosmic volume expansion is realized only in some cases
when the size of the perturbed region is comparable to the horizon
radius of the EdS universe. 

From the above studies, it seems difficult, therefore, that the
accelerated expansion is caused not only by the averaging and
backreaction of sub-horizon perturbations, but also by those of
super-horizon perturbations.

Wiltshire\cite{w1,w2,w3} studied the difference of gravitational
energies and their influence to the clock rates in the different
regions such as the averagedly expanding region, voids and walls
(finite infinity), and considers that this difference brings the observed
accelerated expansion.  Here we take notice of the definition of the
gravitational energy and why it can be so large as to change
remarkably the 
clock rates in different regions.        

\subsection{Fitting}
Without averaging and backreaction of inhomogeneous models, we can
consider the accelerating effect of inhomogeneities by calculating
directly the observational quantities such as the angular diameter
distance and the redshift in inhomogeneous models and comparing them
with the 
counterparts in FLRW models with nonzero $\Lambda$. That is, by
fitting these two models, we can obtain an effective $\Lambda$.

Vanderveld et al.\cite{vand2} studied this fitting by using the 
post-Newtonian
approximation and assuming the linear perturbations which are
normalized with respect to CMB temperature anisotropies and cluster
statistics. Then they obtained an important result that the 
effective $\Lambda$ is
about $0.004$, which is very small, compared with the value required
by the realization of the observed acceleration.

Recently, Marra et al.\cite{marra1,marra2} studied the fitting by
assuming nonlinear perturbations whose amplitudes are much larger than
those expected from the CMB normalization and which are given using
many arranged swiss-cheese models on scales of several hundred
Mpc. From these models they could obtain the effective $\Lambda$ which
is comparable with the value necessary for the observed
acceleration. At present, however, we do not know how these
perturbations with such high amplitudes can exist, without
contradicting with the observed CMB temperature anisopropies and the
other obsevations.

\section{Concluding remarks}
\label{sec:level4} 
Various inhomogeneous cosmological models with a local void have so
far been proposed to explain the observed accelerating property in the
$m$-$z$ relation of SNIa, without a cosmological constant, and their
geometrical structure of the models obtained using LTB solutions
also has been studied in details. At present, on the other hand, there
is no data for SNIa with $z > 1.7$. We should notice that this is very 
important for the model selection, especially to clarify which of 
the concordant model or local-void models are better with respect to 
this observation.

Recently the studies of their consistency with observations such as BAO
(especially RBAS), kinematic Sunyaev-Zeldovich effect, and spectral
distortion at the reionization stage have shown that many models
with a local void are ruled out, and we find that only Gps models with the
narrow range of parameters remain to be examined. So at present the
possibility for these models to survive is small, but may not be zero.

Moreover, the averaging and backreaction of inhomogeneous models and
the fitting with nonzero $\Lambda$ models have been studied. At
present, however, it seems difficult to obtain accelerated expansion
and the 
expected effective $\Lambda$, unless we assume perturbations with
amplitudes much larger than those corresponding to the CMB
normalization or high amplitudes of gravitational energies included
within structures like clusters and voids.    

In these analyses, we have assumed the models based on the Einstein
gravitational theory and the existence of an inflationary early stage.  
If the models should be derived, however, from the other gravitational
theories, such as the superstring theory, the cosmological situation
may be quite different, because we do not know how the inflation
arises, what inflation we have, and whether the expected cosmological
constant can exist or not. Then the inhomogeneous models with a local void 
may play some more role to explain the observed accelerating behavior.

\begin{acknowledgments}
I would like to thank H. Kodama and A. Ishibashi for their kind
organization of the KEK Cosmophysics Workshop DE2008 on ``Is our
Universe really undergoing an accelerating expansion ?'' held in KEK
(Institute of High Energy Physics, Tukuba, Japan) in Dec. 8-12, 2008.
I am grateful to the participants A. Starobinski, R.A. Vandervelt,
A. Notari, D. Wiltshire, M. Kasai, K. Nakao, Y. Nambu, J. Soda and
K.T. Inoue for their helpful discussions.  
\end{acknowledgments}

\end{document}